\documentclass[twocolumn]{aastex63}

\def\simlt{\mathrel{\rlap{\lower 3pt\hbox{$\sim$}}
        \raise 2.0pt\hbox{$<$}}}
\def\simgt{\mathrel{\rlap{\lower 3pt\hbox{$\sim$}}
        \raise 2.0pt\hbox{$>$}}}

\received{? ?, 2023}
\revised{\today} 
\accepted{\today}

\submitjournal{ApJ}

\shorttitle{High amplitude optical variations in Swift~J1910.2$-$0546}
\shortauthors{Saikia et al. 2023}

\usepackage{multirow}

\begin{document}

\title{Seven reflares, a mini-outburst and an outburst:\\
High amplitude optical variations in the black hole X-ray binary Swift~J1910.2$-$0546}

\correspondingauthor{Payaswini Saikia}
\email{ps164@nyu.edu}

\author[0000-0002-5319-6620]{Payaswini Saikia}
\affiliation{Center for Astro, Particle and Planetary Physics, New York University Abu Dhabi, PO Box 129188, Abu Dhabi, UAE \\}

\author[0000-0002-3500-631X]{David M. Russell}
\affiliation{Center for Astro, Particle and Planetary Physics, New York University Abu Dhabi, PO Box 129188, Abu Dhabi, UAE \\}

\author{Saarah F. Pirbhoy}
\affiliation{Center for Astro, Particle and Planetary Physics, New York University Abu Dhabi, PO Box 129188, Abu Dhabi, UAE \\}

\author[0000-0003-1285-4057]{M. C. Baglio}
\affiliation{Center for Astro, Particle and Planetary Physics, New York University Abu Dhabi, PO Box 129188, Abu Dhabi, UAE \\}
\affiliation{INAF, Osservatorio Astronomico di Brera, Via E. Bianchi 46, I-23807 Merate (LC), Italy\\}

\author[0000-0002-1583-6519]{D. M. Bramich}
\affiliation{Center for Astro, Particle and Planetary Physics, New York University Abu Dhabi, PO Box 129188, Abu Dhabi, UAE \\}
\affiliation{Division of Engineering, New York University Abu Dhabi, PO Box 129188, Saadiyat Island, Abu Dhabi, UAE\\}

\author[0000-0003-0168-9906]{Kevin Alabarta}
\affiliation{Center for Astro, Particle and Planetary Physics, New York University Abu Dhabi, PO Box 129188, Abu Dhabi, UAE \\}

\author[0000-0003-3352-2334]{Fraser Lewis}
\affiliation{Faulkes Telescope Project, School of Physics and Astronomy, Cardiff University, The Parade, Cardiff, CF24 3AA, Wales, UK\\}
\affiliation{Astrophysics Research Institute, Liverpool John Moores University, 146 Brownlow Hill, Liverpool L3 5RF, UK\\}

\author{Phil Charles}
\affiliation{Department of Physics \& Astronomy, University of Southampton, Southampton SO17 1BJ, UK\\}

\begin{abstract}

We present long-term (2012--2022) optical monitoring of the candidate black hole X-ray binary Swift~J1910.2$-$0546 with the Faulkes Telescopes and Las Cumbres Observatory (LCO) network. Following its initial bright 2012 outburst, we find that the source displayed a series of at least 7 quasi-periodic, high amplitude ($\sim 3$ mags) optical reflares in 2013, with a recurrence time increasing from $\sim 42$ days to $\sim 49$ days. In 2014, the source experienced a mini-outburst with two peaks in the optical. We also study the recent 2022 outburst of the source at optical wavelengths, and perform a comparative analysis with the earlier rebrightenings. A single X-ray detection and only two radio detections were obtained during the 2013 reflaring period, and only optical detections were acquired in 2014. During the reflaring in both 2013 and 2014, the source showed bluer-when-brighter behavior, having optical colors consistent with a blackbody heating and cooling between 4500 and 9500 K, i.e. the temperature range in which hydrogen starts to ionize. Finally, we compare the flaring behavior of the source to re-brightening events in other X-ray binaries. We show that the repeated reflarings of Swift~J1910.2$-$0546 are highly unusual, and propose that they arise from a sequence of repetitive heating and cooling front reflections travelling through the accretion disk.

\end{abstract}


\section{Introduction} \label{sec:intro}

Low mass X-ray binaries (LMXBs) are systems in which a low-mass ($M<M_{\odot}$) companion star is orbiting a compact object, either a neutron star (NS) or a black hole (BH). The companion (main sequence or evolved) star fills its Roche lobe and transfers mass towards the compact object, forming an accretion disk around it. When this accretion disk becomes unstable, the inflowing matter in the disk heats up, becomes ionized, and this initiates an outburst, in which the optical and X-ray luminosity can increase by several orders of magnitude \citep[e.g.][]{Dubus2001,Lasota2001}. LMXBs emit most of their power in the X-ray band by releasing the gravitational potential energy of the accreted matter. Often during an outburst, collimated synchrotron-emitting compact jets are launched \citep[e.g.][]{corbel2000,Fender04}, analogous to the jets observed in supermassive BHs hosted by active galactic nuclei \citep[e.g.][]{Blandford79}. Accreting BHs, spanning several orders of magnitude in BH mass, follow a correlation between the X-ray and radio luminosity normalized by mass, indicating a coupling between the jet and the inflowing matter \citep[e.g.][]{merloni2003,Falcke2004,saikia2015,saikia2018}.

Outbursts of black hole X-ray binaries (BHXBs) typically last months--years and are quite often characterized by a fast-rise, exponential decay light curve profile \citep[e.g.][and references therein]{Chen1997,Tetarenko2016WATCHDOG}. However, there are many exceptions, with some sources rising slowly, some having multiple peaks, and some displaying flares, dips, plateaus and re-brightenings \citep[e.g.][and references therein]{Buxton2012,Kalemci2013,Zhang2019}. While reflares during outburst decays are fairly common, re-brightenings after the outburst's end, when the source has reached quiescence, have been reported in far fewer sources. These re-brightenings usually peak at a fainter luminosity than the first outburst and last a shorter time; such events are coined mini-outbursts \citep[see][for classifications of re-brightening events in LMXBs]{Zhang2019}. The origin of reflares, and mini-outbursts in particular, are a matter of debate. X-rays from the main outburst heat the companion star, which could increase mass transfer into the disk, causing outburst `echoes' \citep[e.g.][]{Dubus2001,Kalemci2014}. Sometimes these reflares can also be observed in optical and infrared \citep[e.g.][]{Zhang2019}, and they can also be caused by the reactivation of jets during outburst decays \citep[e.g.][]{Jain2001,Kalemci2013,Russell2020}. In neutron star LMXBs, multiple flares (at different timescales) could be caused by the propeller effect, which has been proposed to change the mass accretion rate due to the rapidly rotating neutron star magnetosphere \citep{Hartman2011,Patruno2016_1808}.\\

Historically, many mini-outbursts and late re-brightening events may have been missed, due to their faintness and a lack of either sensitive X-ray telescopes or regular optical monitoring. Long-term optical monitoring of LMXBs, in particular using robotic telescopes, provides an inexpensive way to monitor their activity at low accretion rates, even for long periods of quiescence. Several LMXB outbursts and re-brightening events have been identified using this method \citep[e.g.][]{Callanan1995,CorralSantana2010,Lewis2010,MacDonald2014}, especially more recently \citep[e.g.][]{RussellAlQasim2018,Zhang2019,Pirbhoy2020,Goodwin2020,saikia2021cen,Baglio_submitted,jwaher}. Optical transient surveys have also detected some LMXB brightenings in recent years \citep[e.g.][]{Drake2017,Tucker2018,vanVelzen2019}.

\subsection{Swift~J1910.2$-$0546} \label{sec:j1910}

Swift~J1910.2$-$0546 (MAXI~J1910$-$057, hereafter J1910.2) was independently discovered by the Neil Gehrels Swift Observatory \citep[\emph{Swift};][]{burrows} and the Monitor of All-sky X-ray Image \citep[MAXI;][]{maxi}, when the source went into an outburst in 2012 May \citep{Krimm2012,Usui2012}. The 2012 outburst was extensively studied using X-ray spectral and timing analysis \citep[e.g.]{Nakahira14,Degenaar2014}, optical photometry (Saikia et al. submitted) and spectroscopy \citep{Charles2012,Casares2012}. From these detailed studies, J1910.2 was was found to be a likely BH candidate at a distance of $d > 1.7$ kpc \citep{Nakahira14}. Optical variability of the source suggests the orbital period to be fairly short \citep[$\sim$2-4 hrs,][]{llyod} with an upper limit of $\le 6$~hrs (Saikia et al. submitted), although we note that a larger value is expected from spectroscopic studies \citep[$\ge 6.2$ hrs][]{Casares2012}. 

Following the 2012 outburst, \emph{Swift} and MAXI continued to detect J1910.2 until 2013 Jan, after which the flux levels of the source had decreased below the detection limits. Radio detections were obtained on 2013 Mar 9 and May 3, along with \emph{Swift} observations on Mar 9 (optical detection, X-ray non-detection) and May 10 \citep[X-ray detection,][]{Tomsick2013ATel5063}. No further observations of J1910.2 have been reported since 2013 May, except optical (2015 July) and NIR (2017 April) detections in quiescence \citep{Lopez2019}, until the recent enhancement of activity of the source in 2022. A new X-ray outburst from J1910.2 was detected in 2022 Feb \citep{2022xray}, when it was also found to be prominent in the radio \citep{2022radio} and optical \citep{2022opt1,2022opt2}. The source quickly and steadily decayed at all wavelengths, and was found to be back in optical quiescence by the end of 2022 March \citep{saikiaatel}.

Here we report the long-term optical monitoring of J1910.2 with the Faulkes Telescopes\footnote{\url{http://www.faulkes-telescope.com/}} and Las Cumbres Observatory (LCO)\footnote{\url{https://lco.global/}} network of telescopes from 2012 to 2022. We mainly focus on two periods of activity that were previously undocumented -- a series of strong flaring in 2013, and a faint mini-outburst in 2014. We combine our optical data with \emph{Swift} and MAXI monitoring (in UV and X-ray wavelengths) and radio data from the literature to discuss the optical emission processes in J1910.2 throughout quiescence and outbursts; and explore the various physical explanations behind the flaring activity and the mini-outburst. The observations are described in Section \ref{sec:obs}, and the results are presented and discussed in Section \ref{sec:results}. We include a comparative analysis of the reflares with other BHXB systems in Section \ref{sec:otherLMXBs}, and a summary is provided in Section \ref{sec:conclusions}.

\section{Observations} \label{sec:obs}

\subsection{Faulkes Telescope / LCO monitoring} \label{sec:obsLCO}

We have been monitoring J1910.2 at optical wavelengths since its discovery in 2012, using the 2-m Faulkes Telescopes at Haleakala Observatory (Maui, Hawai`i, USA) and Siding Spring Observatory (Australia), as well as the 1-m telescopes at Siding Spring Observatory (Australia), Cerro Tololo Inter-American Observatory (Chile), McDonald Observatory (Texas), Teide Observatory (Tenerife) and the South African Astronomical Observatory (SAAO, South Africa) of the LCO network \citep[][]{Brown2013}. The observations were performed in the Bessell $B$, $V$, $R$ and SDSS $i^{'}$ filters, as part of an on-going monitoring campaign of $\sim$50 LMXBs \citep{Lewis2008}. We use the ``X-ray Binary New Early Warning System (XB-NEWS)'' data analysis pipeline \citep{Russell2019,Pirbhoy2020,Goodwin2020} for calibrating the data and performing aperture photometry (see Saikia et al. submitted, for more details). This process resulted in photometric measurements of J1910.2 in a total of 123 ($B$), 74 ($V$), 85 ($R$) and 211 ($i^{'}$) images between 2012 June 14 (MJD 56092) and 2022 Mar 20 (MJD 59658).

We note that J1910.2 lies in the Galactic plane, with a few faint stars within 2$^{\prime\prime}$ of the source position \citep{Lopez2019}. These stars may contribute to the quiescent flux measurements, but are too faint to affect the active interval photometry. Due to the limitation in the resolution and sensitivity of the Faulkes and LCO Telescopes, it is difficult to provide a proper numerical estimate of the contribution of the two neighbouring stars to the quiescent magnitude of the source.

\subsection{Archival X-ray and UV monitoring} \label{sec:obsXray}

We acquired the X-ray detections of J1910.2 obtained with the X-Ray Telescope \citep[XRT;][]{burrows}  onboard \emph{Swift}, using the online \emph{Swift}/XRT data products generator\footnote{\url{https://www.swift.ac.uk/user objects/}} maintained by the \emph{Swift} data center at the University of Leicester \citep[see][]{evans1,evans2}. The source was observed for 67 days during its 2012 outburst (see Saikia et al. submitted) in the Windowed Timing mode \citep[WT,][]{hill}. Due to Sun constraint, no observations were taken by \emph{Swift}/XRT from 2012 November until 2013 March. It was again observed in Photon Counting mode \citep[PC,][]{hill} for 5 days between 2013 March to September, with exposures ranging from $\sim$1000 s to $\sim$2000 s (Observation ID 00032742), and was detected only once. In addition to the \emph{Swift}/XRT light curve, we also acquired the 2--10 keV MAXI/GSC light curve \footnote{http://maxi.riken.jp/top/index.html}. Unfortunately, during the flaring activity of J1910.2, MAXI only detected the system once above 3$\sigma$ significance.

We also retrieved the publicly available \emph{Swift} Ultraviolet and Optical Telescope \citep[UVOT;][]{Roming} observations from the NASA/HEASARC data center. We use the pipeline processed images and follow the {\tt uvotsource} HEASOFT routine to obtain the magnitudes of the source using an aperture size of 5$^{\prime\prime}$ centered on the source. During the 2012 outburst, the source was detected in almost all the epochs observed by \emph{Swift}/UVOT, for a varying range of exposures between $\sim$20-1000 s (see Saikia et al. submitted). However, most of the observations during the flaring period and the faint mini-outburst during 2013 and 2014 were non-detections (the significance of the detection above the sky background is lower than 5$\sigma$), despite having much longer exposure times (even for $\sim$1000 s exposures).

\begin{figure*}
\centering
\includegraphics[height=7.2cm,angle=0]{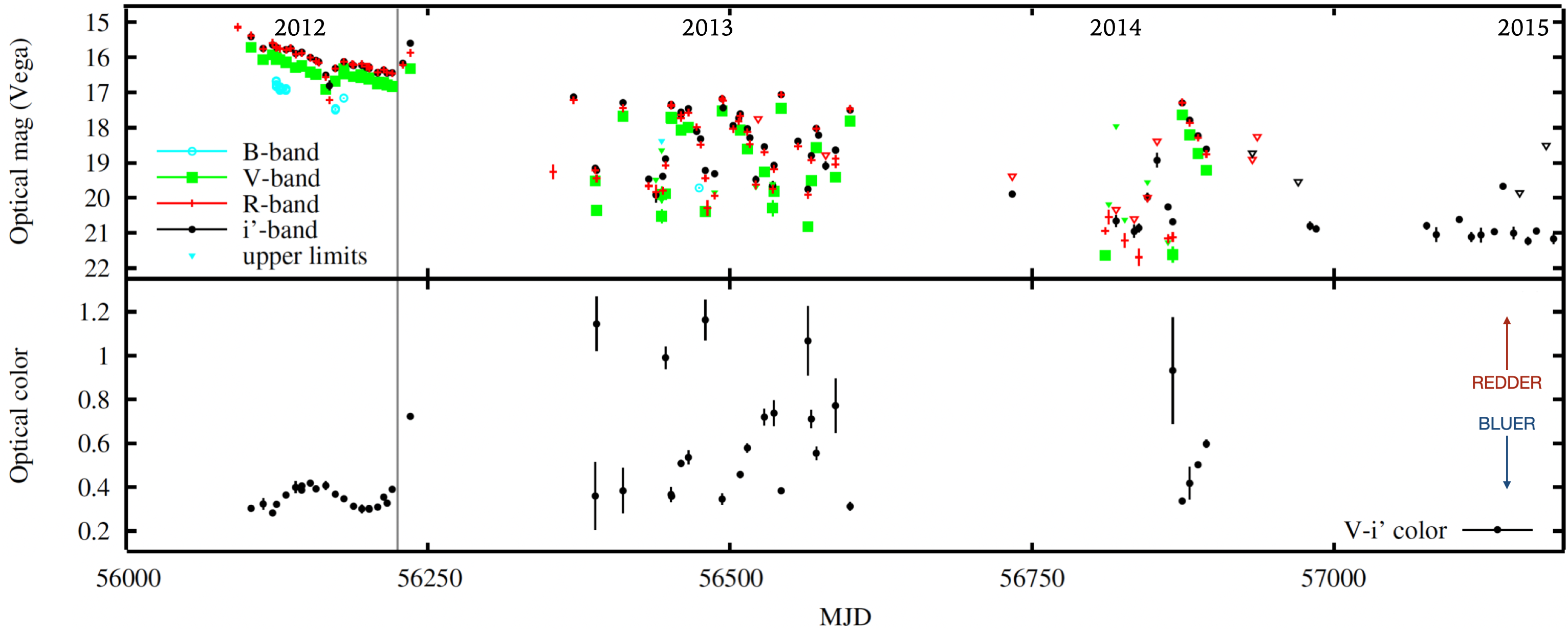}
\caption{Long-term light curve (top; in $B$, $V$, $R$ and ${i}^{\prime }$) and colour (bottom; $V$-${i}^{\prime }$) of J1910.2 from 2012 to 2015. The grey vertical line (at MJD 56225) shows the transition to a pure hard state during the 2012 outburst. }
\label{fig:lc-large}
\end{figure*}

\subsection{Radio data} \label{sec:obsRadio}

We searched the literature for detections of J1910.2 after the 2012 outburst. In 2013, detections were acquired by the Australia Telescope Compact Array (ATCA) in March and May at 5.5 and 9 GHz on both dates, with average flux densities of 0.06 mJy in March and 0.3--0.4 mJy in May \citep[no errors are given;][]{Tomsick2013ATel5063}. It was again detected during its 2022 outburst with the Arcminute Microkelvin Imager Large Array \citep[AMI--LA;][]{amizwart,ami2} at 15.5 GHz \citep{2022radio}.

\section{Results and Discussion} \label{sec:results}

\begin{table}
\centering
\caption{Summary of optical periods of activity in J1910.2 during 2012-2022.}
\begin{tabular}{ l l l l}
\hline
\hline
Activity$^1$ & Year  & N$_{\rm fl}$$^2$ & Peak $i^{'}$(mag) \\
\hline
Outburst	&  2012 May -- 2013 Jan	 & 1 &	 15.41$\pm$0.01 \\
Reflares	&  2013 Feb -- Nov	 & $\geq 7$ &	 $\sim$17.0-17.5 \\
Mini-outburst	&  2014 June -- Sept	 & 2 &	 17.30$\pm$0.01 \\
Outburst	&  2022 Feb -- Mar	 & 1 &	 16.45$\pm$0.01 \\
\hline
\end{tabular}
$^1$Re-brightening classification based on \cite{Zhang2019}.
$^2$Number of flares seen during the period of re-brightening.
\par
\end{table}

In Fig. \ref{fig:lc-large} we present the long--term LCO optical data of J1910.2 in $B$, $V$, $R$ and ${i}^{\prime }$ as well as the $V$-${i}^{\prime }$ color, from the start of the 2012 outburst until 2015. After the main 2012 outburst there is a gap (Sun contraint), following which, in 2013, J1910.2 was found to be undergoing high amplitude flaring (see section \ref{sec:2013reflares}).  In 2014 there was a short mini-outburst (section \ref{sec:2014outburst}) followed by quiescence.  The colour variability (Fig. 1) shows that, during the reflares and mini-outburst, the source follows a bluer-when-brighter behaviour.

Since 2015 the source has remained in quiescence, as far as our monitoring can tell, until 2022 Feb, when it was observed to undergo a new outburst (see Section 3.4). Table 1 summarises these periods of optical activity in J1910.2 during 2012-2022.

\subsection{The 2013 reflares} \label{sec:2013reflares}

\begin{figure}
\centering
\includegraphics[width=8.0cm,angle=0]{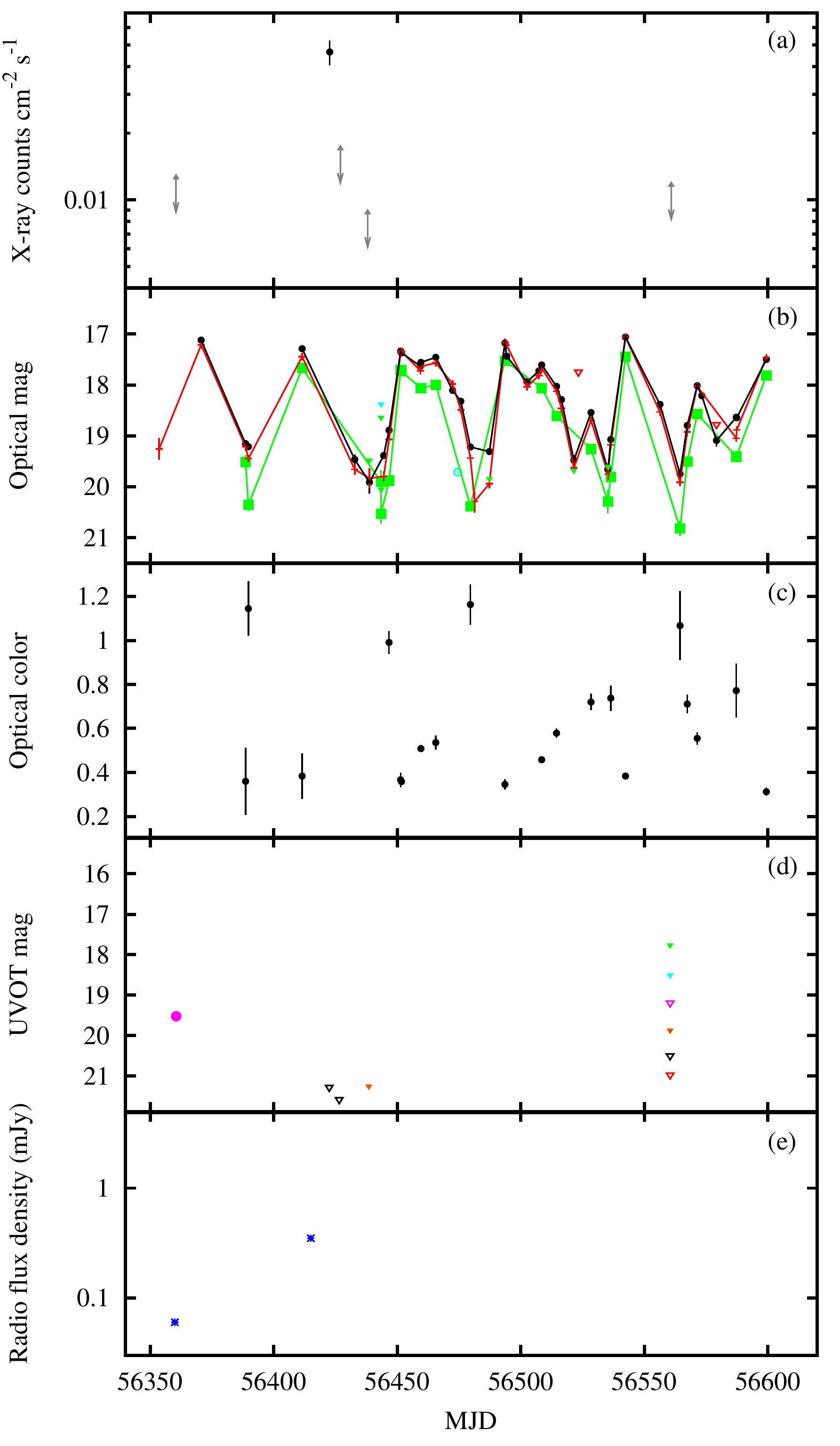}
\includegraphics[width=5.7cm,angle=270]{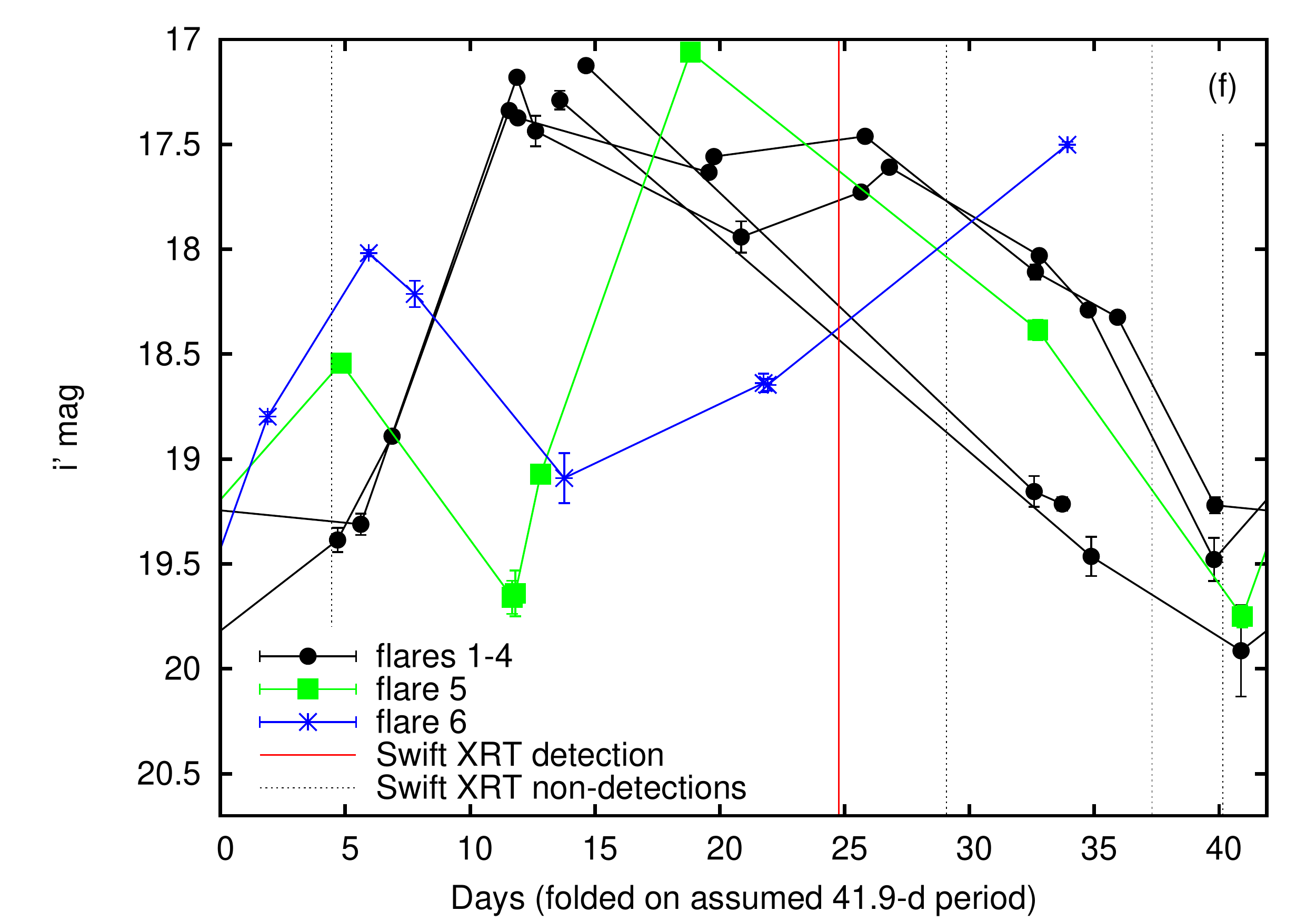}
\caption{(Upper box) Flaring behaviour of J1910.2 during the 2013 reflares (see section 3.1).  (a) Swift/XRT light curve; (b) Optical magnitudes in $V$ (green squares), $R$ (red plus), $B$ (blue circles) and ${i}^{\prime }$ (black dots), with upper limits shown as inverted triangles; (c) optical colour ($V$-${i}^{\prime }$); (d) Swift UVOT magnitudes in $u$ (magenta circles) and upper limits shown as inverted triangles in $v$ (green), $b$ (cyan), $u$ (magenta), $uvw1$ (orange), $uvw2$ (red) and $uvm2$ (black); (e) radio flux density (mJy). (Lower box) (f) 2013 reflares folded on P=41.9d (the interval between the fast rises of flares 3 and 4)}
\label{fig:lc-2013}
\end{figure}

From MAXI and Swift X-ray data, \cite{Nakahira14} report that the 2012 outburst of J1910.2 ended around 2013 Jan 26 (MJD 56318). Due to Sun constraints we have no optical coverage during the 2012 outburst decay, and when monitoring was resumed in 2013 Mar we found J1910.2 in a flaring state.  It displayed at least 7, high amplitude ($\sim$3  magnitude), quasi-periodic optical reflares (with the interval between reflares increasing from $\sim$42 to $\sim$49 days), which continued for at least 8 months. Due to the lack of coverage before 2013 Mar, it is not evident if and when the source entered quiescence before the re-brightening, and it is also not possible to constrain the exact date when the reflaring started or ended.

We plot the 2013 re-brightening activities of the source in Fig. \ref{fig:lc-2013}. During the interval of 2013 Feb 27 to Nov 4 (MJD 56350 to MJD 56600), the source had unusually extreme reflares in all optical bands, which had not been seen before. The optical magnitude during the peak of these reflares reached ${i}^{\prime }\sim$17-17.5. Between any two consecutive flares, the magnitudes did not drop to the quiescent value, and remained around ${i}^{\prime }\sim$19-20. For a rough comparison, J1910.2 is found to have a quiescent ${i}^{\prime }$ of 22.18$\pm$0.04 using the William Herschel Telescope with the auxiliary port camera \citep[ACAM, 2015 July 19,][]{Lopez2019}. The optical color ($V$-${i}^{\prime }$) roughly decreased during the rise of the flare, following a bluer-when-brighter behavior (see Fig 2c, Fig 7). It was seen to be the lowest during the peak of the flares, and the color reddened during the decay of the flares. Radio and X-ray observations carried out in this period with \emph{ATCA} (2013 Mar 9 and May 3) and \emph{Swift}/XRT (2013 Mar 9 and May 10) show that the source was probably flaring in these bands as well \citep{Tomsick2013ATel5063}. While in 2013 Mar, the authors report a radio flux of 0.06 mJy, it increased to 0.3-0.4 mJy in 2013 May. On the other hand, the source was not detected above 3$\sigma$ significance with Swift/XRT (0.6--10 keV) during the 2013 March observation, but it was observed to be brighter in 2013 May. Inspecting the MAXI light curves in the energy band 2.0--10.0 keV for the same period, we found that J1910.2 was only detected once (MJD 56451) above 3$\sigma$ significance. Taking into account the \textit{Swift}/XRT and MAXI detections, we can estimate a lower and upper limit for the X-ray flux in the energy band 2.0--10.0 keV, being these $\sim 4.0\times 10^{-12}$ erg cm$^{-2}$ s$^{-1}$ and $\sim 3.0\times 10^{-10}$ erg cm$^{-2}$ s$^{-1}$, respectively.

The average optical cycle time of the reflares increases with time from $\sim$42 to $\sim$49 days. In Fig. \ref{fig:lc-2013}(f) we show the first six reflares folded on a period of 41.9 d (the time between the fast rises of flares 3 and 4). As evident from the figure, the four initial reflares have a very similar duration consistent with $\sim 41.9$ d. The first two reflares (flares 1 and 2) have a fairly sharp peak before decaying, with comparable rise and decay time. The next two (flares 3 and 4) have a rise time similar to the previous ones, but show an extended peak lasting $\sim 15$ d, before the decay. For all these first four reflares, the rise time from the minimum to the peak of the reflare is $\sim 6$ d. The last two flares (flares 5 and 6) have a double-peaked morphology, with the first peak being faint and the second peak being a similar magnitude to the first 4 flares, and a slightly longer period (seen in Fig. \ref{fig:lc-2013}(f) as a delay in the bright peak for these flares). Multiple reflares displaying such periodic behaviour have been previously observed in many dwarf novae \citep[see e.g.][]{kato}, but it is rarely seen in LMXBs (see Section \ref{sec:otherLMXBs} for a detailed discussion). \\

\begin{figure}
\centering
\includegraphics[width=8.5cm,angle=0]{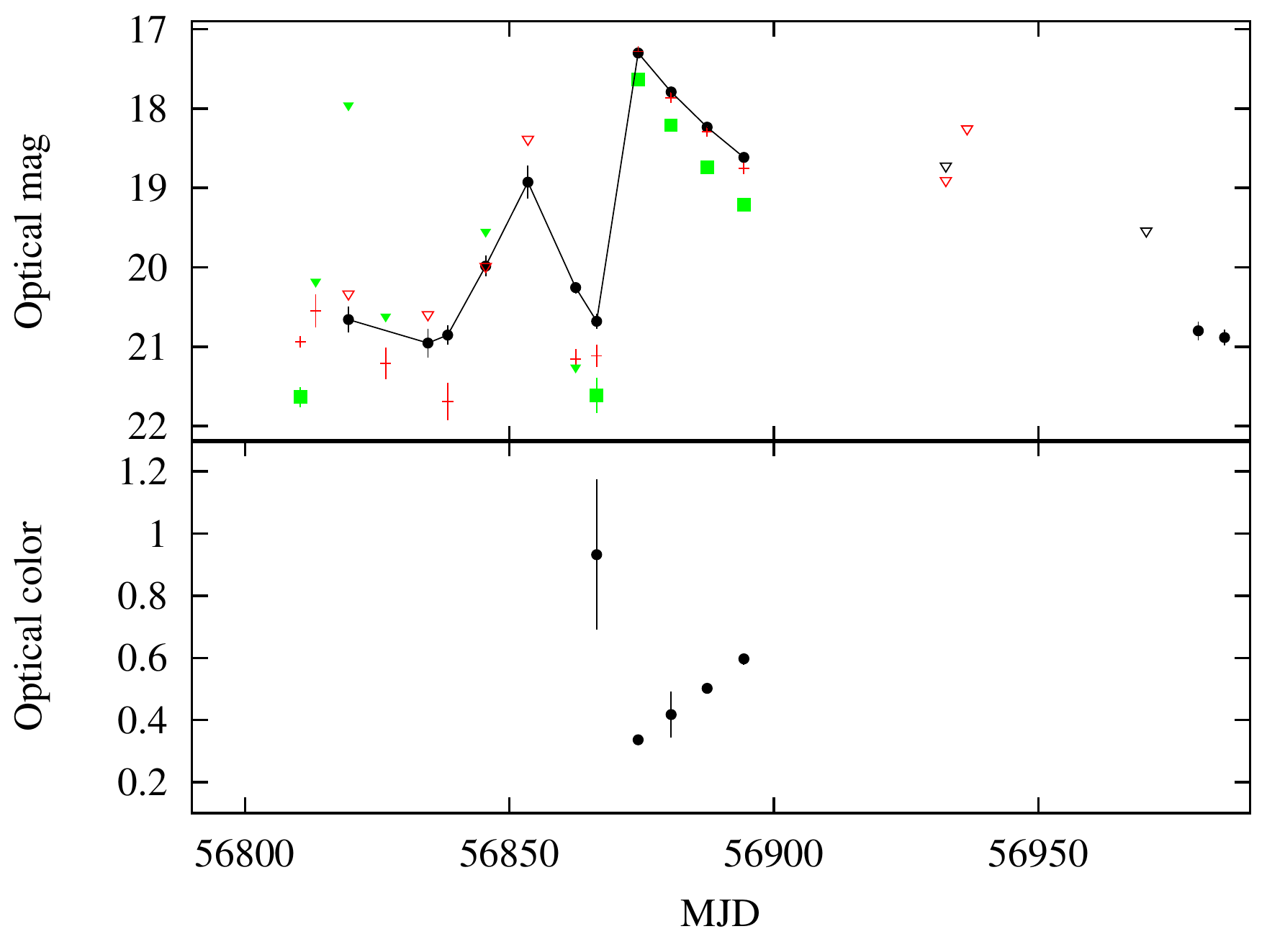}
\caption{2014 mini-outburst of J1910.2. (Upper) Optical light curve in $V$ (green squares), $R$ (red plus) and ${i}^{\prime }$ (black circles), and upper limits as inverted triangles. The ${i}^{\prime }$ points are joined up to show its evolution. (Bottom) Optical colour ($V$- ${i}^{\prime }$) evolution.}
\label{fig:lc-2014}
\end{figure}

\subsection{The 2014 mini-outburst} \label{sec:2014outburst}

When optical monitoring of J1910.2 was resumed in 2014, the source was found in a variable state close to quiescence (${i}^{\prime }\sim$ 20.7-21.7, see Section 3.9) shortly before getting brighter again (see Fig. \ref{fig:lc-2014}), showing two consecutive peaks on MJD 56853.5 (${i}^{\prime }$ = 18.93$\pm$0.21) and a brighter one on MJD 56874.3 (${i}^{\prime }$ = 17.30$\pm$0.01). There was a single LCO detection of the source between the 2013 reflares and the 2014 re-brightening (2014 Mar 17, MJD 56733.6, ${i}^{\prime }$ =19.89$\pm$0.09), which is much fainter than the reflares, but brighter than the typical quiescence value obtained with LCO (${i}^{\prime }\sim$ 20.7-21.7, see Section 3.9). Due to the lack of continuous observations during that period, it cannot be confirmed if the 2013 reflares were going on for the whole year and the re-brightening events seen in 2014 were just a continuation of the 2013 reflares. 

\begin{figure}
\centering
\includegraphics[width=8.3cm,angle=0]{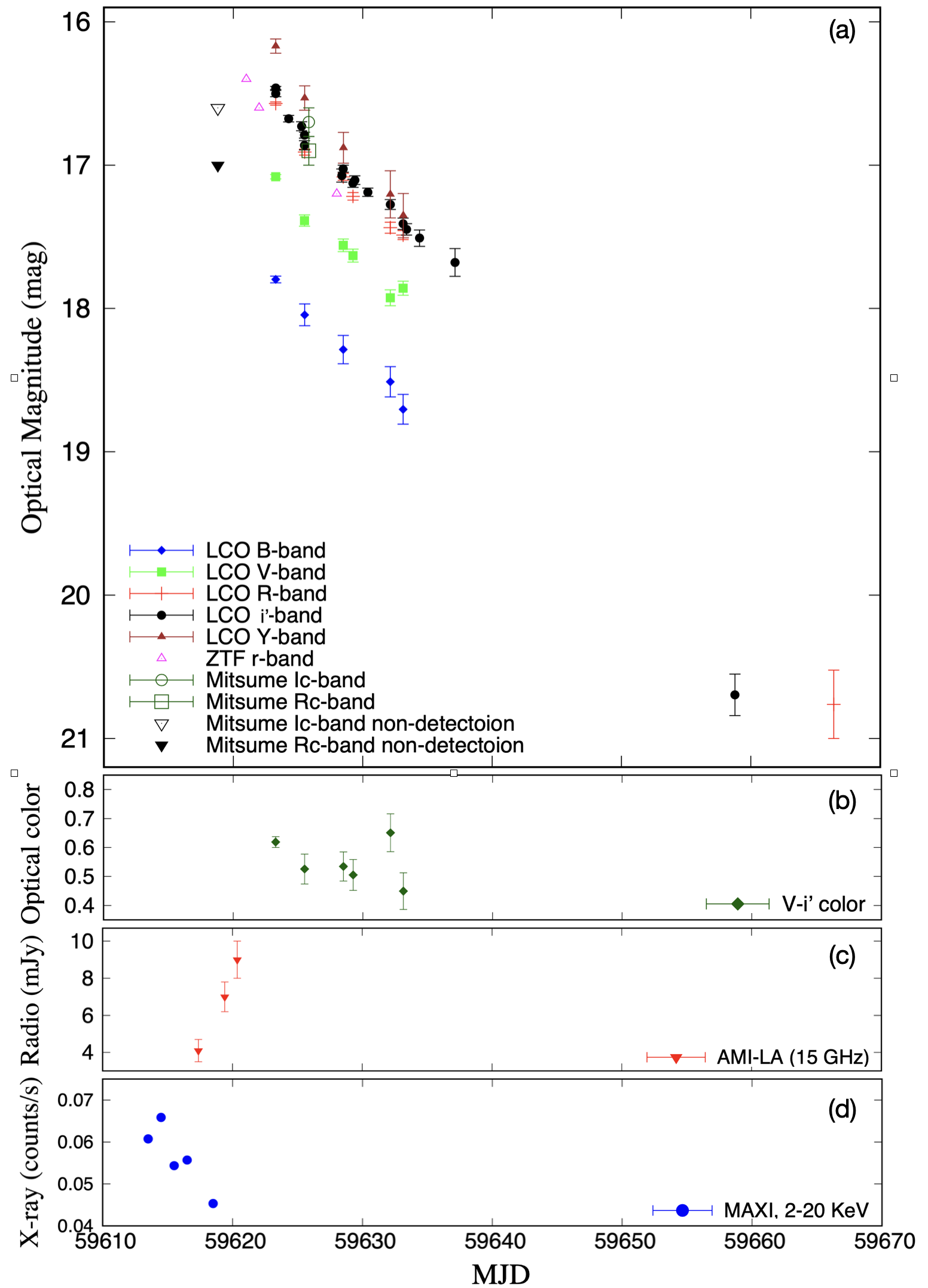}
\caption{2022 outburst of J1910.2. (a) Optical light curve in $B$ (blue, filled diamonds), $V$(green, filled squares), $R$ (red, plus), ${i}^{\prime }$ (black circles) and $Y$ (brown, triangles), and upper limits as inverted triangles.  Also plotted are the available data in the literature from the MITSuME telescope, including non-detections \citep[inverted triangles,][]{2022opt1} and ZTF telescope \cite[magenta, open triangles,][]{2022opt2}. (b) Optical colour ($V$- ${i}^{\prime }$) evolution. (c) Radio flux density (in mJy) obtained with the AMI-LA telescope at 15 GHz \citep{2022radio}. (d) MAXI 2-20 keV daily average light curve (for data with $\geq 4 \sigma$ significance).}
\label{fig:lc-2022}
\end{figure}

The peak of the 2014 re-brightening on May 8 (MJD 56874.3, ${i}^{\prime }$ = 17.30$\pm$0.01) is almost $\sim$2 mags fainter than the 2012 outburst peak on MJD 56103.6 (${i}^{\prime }$ = 15.41$\pm$0.01). Although comparable to the peak magnitudes observed during the 2013 reflares (${i}^{\prime }$-band range $\sim$ 17.0-17.5, see Table 1), we classify the 2014 re-brightening as a mini-outburst because the source had reached close to quiescence before the apparent brightening. Moreover, it follows a typical double-peaked outburst profile with a sudden rise from quiescence followed by an exponential decay after the second peak. The evolution of the optical color ($V$-${i}^{\prime }$) during the mini-outburst also follows a bluer-when-brighter behavior, similar to the 2013 re-flares. This is clearly observed during the second peak of the mini-outburst, where the source is bluest at the peak, and slowly reddens as it decays during the return of the mini-outburst to quiescence.

\subsection{The 2022 outburst} \label{sec:2022outburst}

Recently, renewed X-ray activity was detected in J1910.2 by MAXI/GSC on 2022 Feb 4 (MJD 59614), with the 2-6 keV flux reaching 17 mCrab on Feb 5 (MJD 59615), and then gradually declining to $\sim$7 mCrab on Feb 7 \citep[MJD 59617,][]{2022xray}. The source quickly faded below the detection limit in soft X-rays, and returned to close to quiescence (See Fig. \ref{fig:lc-2022}d). It was detected in the radio by the Arcminute Microkelvin Imager Large Array \citep[AMI--LA;][]{amizwart,ami2} at 15.5 GHz, with integrated fluxes of 4.1$\pm$0.6 mJy on 2022 FeB 7 (MJD 59617.377), 7.0$\pm$0.8 mJy on Feb 9 (MJD 59619.411) and 9.0$\pm$1.0 mJy on Feb 10 (MJD 59620.376), indicating that the source was rapidly brightening \citep[See Fig. \ref{fig:lc-2022}c, values obtained from][]{2022radio}.

LCO first detected J1910.2 during the recent activity on 2022 Feb 13th (MJD 59623.27), after the source came out of the Sun constraint (See Fig. \ref{fig:lc-2022}a). At that time, it was already at the peak, or at the early decline stage of the outburst (with ${i}^{\prime } \sim$16.45$\pm$0.01). This is brighter than the previous re-brightening events of 2013 (flares with peaks in the range of ${i}^{\prime } \sim$17.0--17.5 mag,) and the mini-outburst of 2014 (${i}^{\prime } \sim$17.30$\pm$0.01), and fainter than the previous 2012 outburst with ${i}^{\prime } \sim$15.41$\pm$0.01 on 2012 June 25 (MJD 56103.6, see table 1). An optical non-detection was reported on 2022 Feb 8 (MJD 59618.85) with the MITSuME 50 cm telescope $Akeno$, implying a 5-sigma upper limit of $R_c>$17.0 and $I_c>$16.6, before brightening and being detected on 2022 Feb 15 (MJD 59625.84) with $R_c$=16.9$\pm$0.1 and $I_c$=16.7$\pm$0.1 \citep{2022opt1}. It was also detected by the Zwicky Transient Facility \citep[ZTF;][]{ztf} on 2022 Feb 11th (MJD 59621) with $r\sim$16.4, which got gradually fainter with $r\sim$16.6 on Feb 12th (MJD 59622) and $r\sim$17.2 on Feb 18 \citep[MJD 59628;][]{2022opt2}. Following the re-brightening classification scheme based on \cite{Zhang2019}, we classify the recent activity as a new outburst, as the flux reached quiescence before the re-brightening event, and the time separating the start of the quiescent period (after the end of the last activity) from the start of the recent re-brightening is much larger than the duration of the outburst.

\begin{figure}
    \centering
   \includegraphics[height=6.0cm,angle=0]{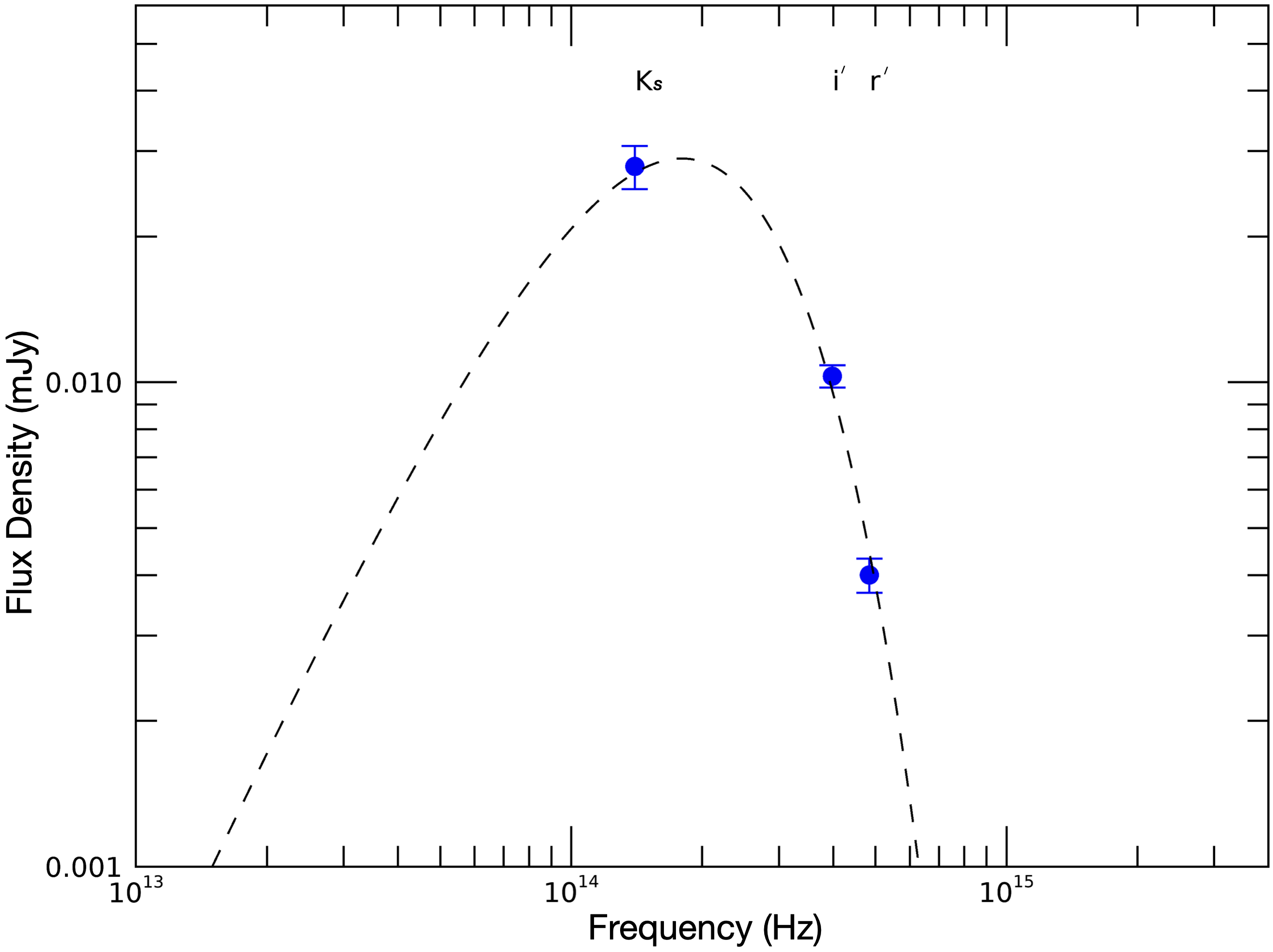}
   \caption{Single temperature blackbody fit to the intrinsic (de-reddened) IR/optical spectrum during quiescence, with optical fluxes obtained from \cite{Lopez2019}.}
   \label{Fig:bbq}
\end{figure}

\begin{figure*}
    \centering
    \includegraphics[height=12.2cm,angle=0]{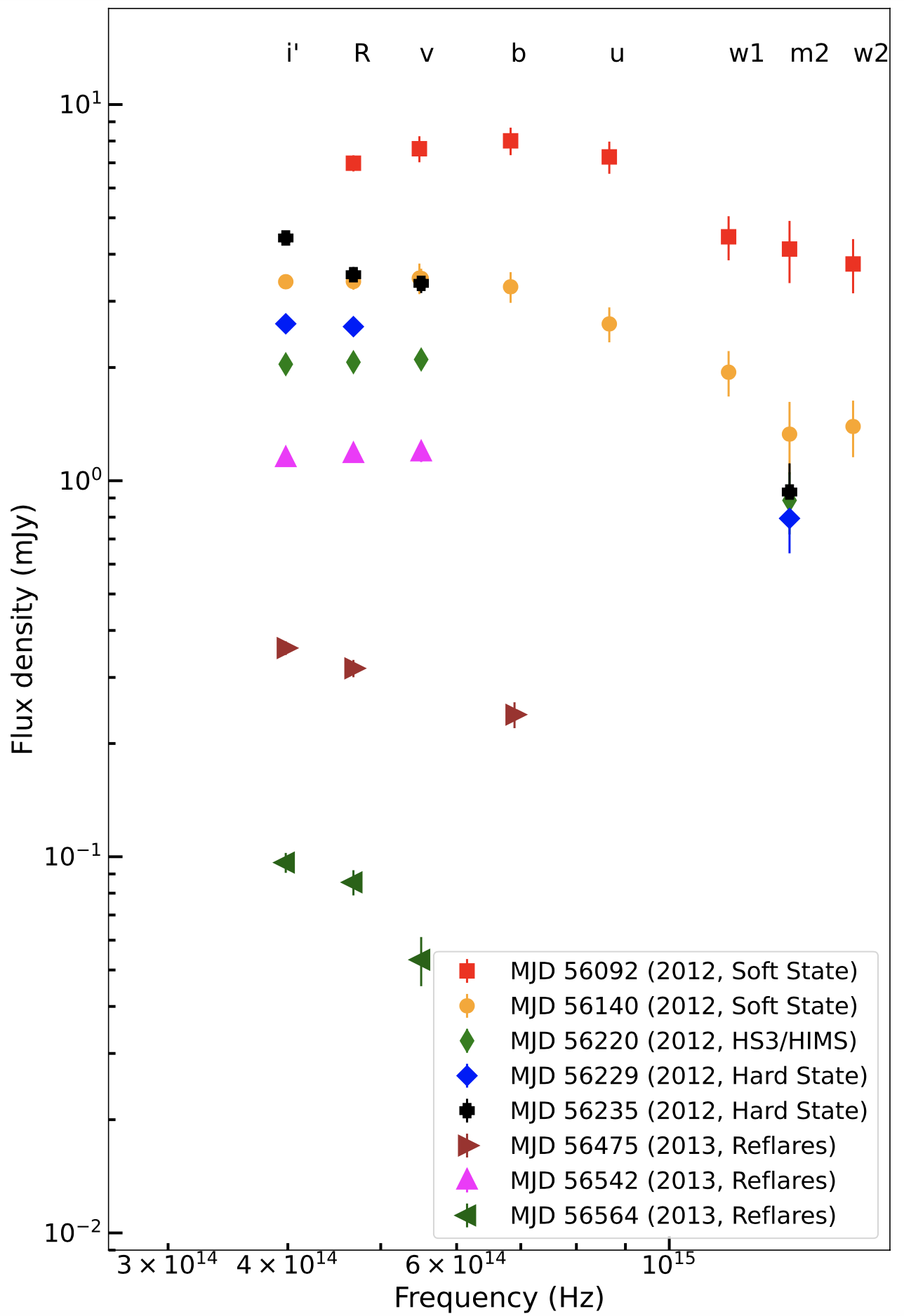}
    \includegraphics[height=12.2cm,angle=0]{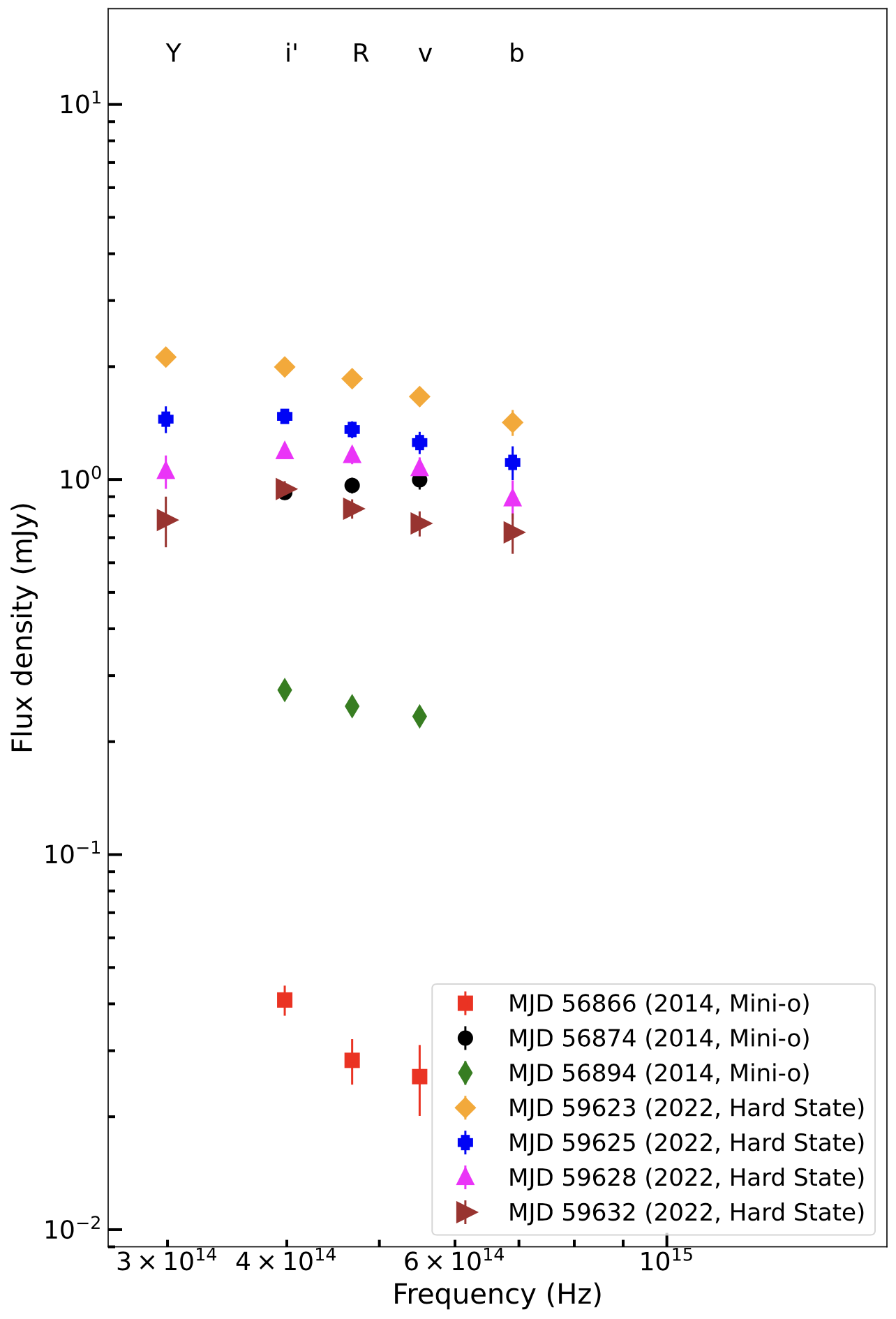}
    \caption{Intrinsic (de-reddened) optical/UV spectra of J1910.2 with quasi-simultaneous data (taken within 24 hours) during the 2012 and 2022 outbursts, as well as the 2013 re-flares and the 2014 mini-outburst.}
    \label{Fig:SEDs}
\end{figure*}

\subsection{Spectral energy distributions} \label{sec:sed}

We build the de-reddened spectra and spectral energy distributions (SEDs) of J1910.2 using quasi-simultaneous observations (within 24 hours) both in quiescence (see Fig. \ref{Fig:bbq}), and in the bright episodes of the 2013 reflares, the 2014 mini-outburst and the 2022 outburst (see Fig. \ref{Fig:SEDs}). We also overplot a few SEDs from different spectral states of its discovery outburst in 2012 for comparison (see Saikia et al. submitted, for the evolution and the naming of the spectral states in J1910.2). De-reddened fluxes were obtained from the calibrated magnitudes using a hydrogen column density value of $N_H=(3.5\pm0.1)\times 10^{21}\, \rm cm^{-2}$ \citep{Degenaar2014} and the \citet{Foight16} and \citet{Cardelli1989} extinction laws to estimate the absorption coefficients (see Table 1 in Saikia et al. submitted, for more details).

For J1910.2, a simple fitting of the quiescent spectrum \citep[using values obtained from][]{Lopez2019} with a single-temperature blackbody gives a value of $\sim$3040 K (Fig. \ref{Fig:bbq}). Assuming that there is no accretion activity at these lowest fluxes, then this temperature is consistent with an M4-type star, of mass $\sim 0.3\,M_{\odot}$ and  radius $\sim\,0.3 R_{\odot}$ (see Saikia et al. submitted).

During the outbursts and re-brightening episodes, the optical/UV spectra are found to be fairly smooth. For the brighter outburst epochs, we find a slightly positive to flat slope in the optical ($\alpha_{R-b}$= -0.04 to 0.35, where $F_{\nu}\propto \nu^{\alpha}$), and a negative slope in the UV ($\alpha_{u-w2}$= -1.0 to -1.2, see Fig 6). The SEDs peak around $V$ or $B$ for the brighter epochs of the 2012 outburst, but appear redder (around ${i}^{\prime }$) for the 2022 outburst. During the re-brightening epochs of 2013 reflares and 2014 mini-outburst, the ${i}^{\prime }$ flux is generally found to be brighter than the higher frequencies, unlike the brighter epochs of the 2012 outburst (see Table 2 for a comparison of the optical spectral indices). This could be a hint of the blackbody peak shifting to lower frequencies, as the luminosity decreases. The overall shape of the optical/UV spectra is consistent with the outer regions of a blue, X-ray irradiated accretion disk \citep[e.g.][]{hynes2005}.

\begin{table}
\centering
\caption{List of optical spectral indices for the spectra presented in Fig. 6 ($\alpha_{{i}^{\prime }-b}$, unless specified).}
\begin{tabular}{ l l l l}
\hline
\hline
Year & MJD  & State & Spectral Index \\
\hline
2012	&  56092	 & Soft &	 $\alpha_{R-b} = $0.35$\pm$0.09 \\
2012	&  56140	 & Soft &	 $\alpha_{{i}^{\prime }-b} = $-0.04$\pm$0.06 \\
2012	&  56220	 & HS3/HIMS &	 $\alpha_{{i}^{\prime }-v} = $0.09$\pm$0.01 \\
2012	&  56229	 & Hard & $\alpha_{{i}^{\prime }-R} \sim $ -0.11$^a$ \\
2012	&  56235	 & Hard &	 $\alpha_{{i}^{\prime }-v} = $-0.85$\pm$0.30 \\
2013	&  56475	 & Reflares & $\alpha_{{i}^{\prime }-b} = $-0.74$\pm$0.01\\
2013	&  56542	 & Reflares & $\alpha_{{i}^{\prime }-v} = $0.11$\pm$0.03 \\
2013	&  56564	 & Reflares & $\alpha_{{i}^{\prime }-v} = $-1.83$\pm$0.63 \\
2014	&  56866	 & Mini-outburst & $\alpha_{{i}^{\prime }-v} = $-1.45$\pm$0.48 \\
2014	&  56874	 & Mini-outburst & $\alpha_{{i}^{\prime }-v} = $0.24$\pm$0.02\\
2014	&  56894	 & Mini-outburst & $\alpha_{{i}^{\prime }-v} = $-0.50$\pm$0.07\\
2022	&  59623	 & Hard & $\alpha_{{i}^{\prime }-b} = $-0.38$\pm$0.08\\
2022	&  59625	 & Hard & $\alpha_{{i}^{\prime }-b} = $-0.23$\pm$0.11\\
2022	&  59628	 & Hard & $\alpha_{{i}^{\prime }-b} = $0.05$\pm$0.16\\
2022	&  59632	 & Hard & $\alpha_{{i}^{\prime }-b} = $-0.04$\pm$0.26\\
\hline
\end{tabular}
\par
$^a$ Only two data-points available, unable to get uncertainity.
\end{table}

\subsection{Color Evolution} \label{sec:colors}

\begin{figure*}
\centering
\includegraphics[height=10.3cm,angle=0]{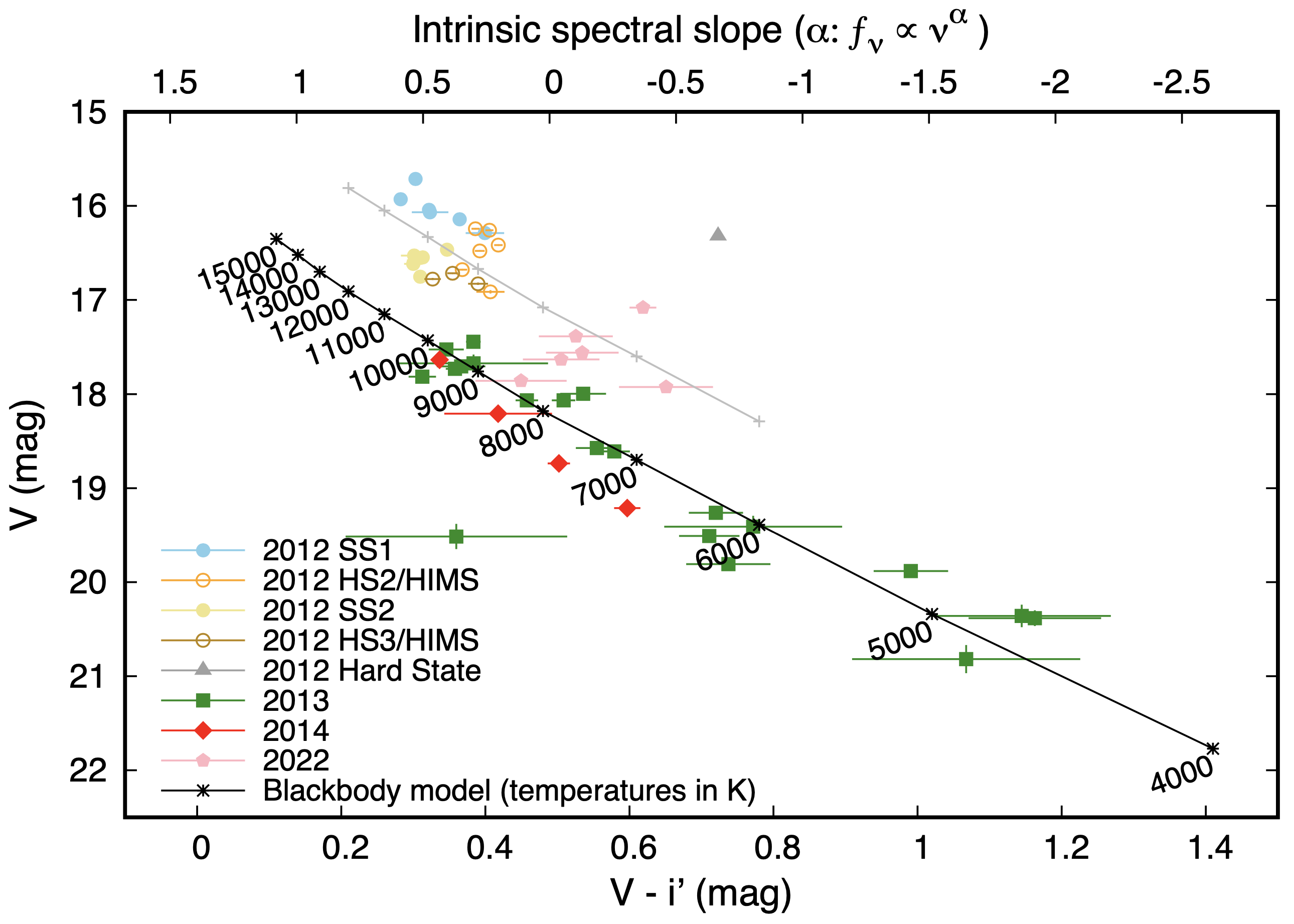}
\includegraphics[height=10.1cm,angle=0]{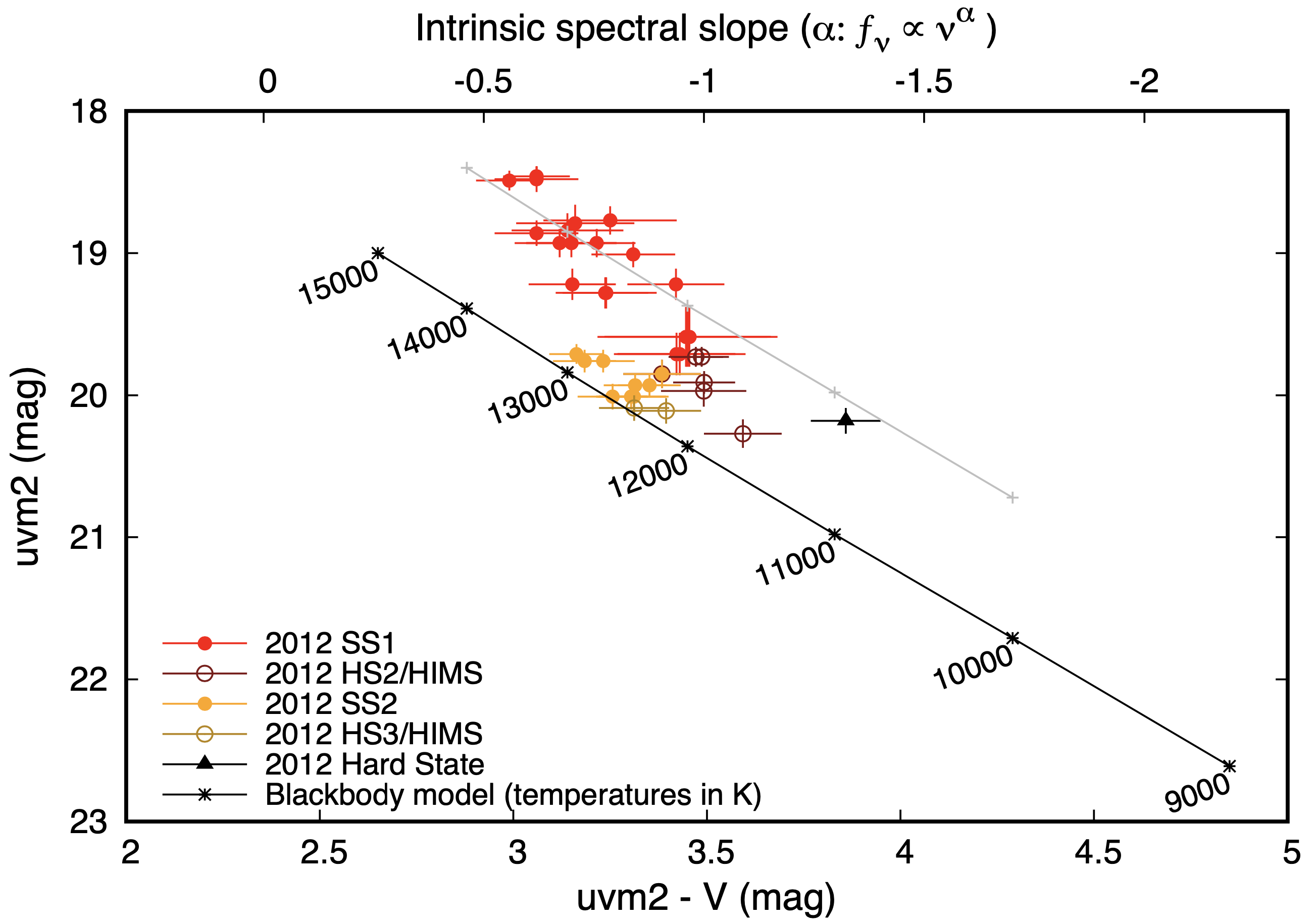}
\caption{Color-magnitude diagrams (CMD) of J1910.2 in (upper) $V$ vs $V$-${i}^{\prime }$, and (lower) $uvm2$ vs $uvm2$-$V$.  The black solid lines show points from single-temperature blackbody models heating up and cooling (with normalisation fixed according to the trend in the $V$ vs $V$-${i}^{\prime }$ CMD for both plots). The grey lines show a different normalisation to better fit only the 2012 outburst.}
\label{fig:CMD}
\end{figure*}

In order to analyse the color evolution of J1910.2, we plot the optical color-magnitude diagram (CMD) using quasi-simultaneous $V$-band and ${i}^{\prime }$-band magnitudes (Fig. \ref{fig:CMD}, top panel), and the optical/UV CMD using $uvm2$-band and $V$-band magnitudes (Fig. \ref{fig:CMD}, bottom panel). The different states of the 2012 outburst, the subsequent re-brightening events, and the 2022 outburst are shown in different colors and symbols to distinguish their temperature ranges and study their comparative behaviours. We also plot the single temperature blackbody model of \cite{Maitra2008}, which approximates the emission of an X-ray irradiated outer accretion disk \citep[see also][]{Russell2011, Zhang2019, baglio2020, Baglio_submitted,saikia1716}. The normalization of the blackbody model depends on various factors, including the accretion disk radius which can be estimated from the system masses, orbital period, inclination, source distance, disk filling factor, disk warping, etc. As many of these parameters are poorly constrained, we fix the normalization value to what best describes the trend in the optical CMD as it has the most amount of data. Amongst the observations, we optimize the normalization so as to cover the widest range of observed optical colour ($V$-${i}^{\prime }$). We use the same normalization also for the optical/UV CMD. 

We find that the observations taken during the reflares in 2013 and the mini-outburst in 2014 are well represented by the disk model (shown as a solid black line in Fig. 7). During these re-brightening events, the outer disk temperature is approximately between 4500 K to 9500 K. This covers the expected temperature range where hydrogen in the disk gets ionized ($\sim$7000-10000K). We find that the disk temperature during these re-brightening events repeatedly increases and then decreases, suggesting that the reflares are caused by the continuous waves of heating and cooling flowing through the accretion disk.

However, the initial outburst of 2012 does not completely follow the same model of a single temperature blackbody heating and cooling. The temperature during this main outburst is higher than the H ionization temperature ($\sim 11,000$ K), and the emission is redder and/or brighter than what is expected from the disk model. We find that the data are better represented by a single temperature blackbody model with a different normalization (shown as a solid grey line in Fig. 7). This could either be due to significant contribution to the optical emission from additional components such as synchrotron emission from a jet, or because the viscous disk starts to dominate, or else because of disk warping. The factor difference between the two normalizations is 2.75 in flux, indicating the expected increase in the surface area needed to explain the brighter points from the initial outburst, compared to the reflares and mini-outburst.

We note the synchrotron emission is unlikely to be the dominant cause, as the brighter and redder trend is also observed in the soft state, when we do not expect the jet to be present. Jets have been observed in the IR/optical during transition from the soft state to the soft-intermediate state \citep[e.g.][]{Russell2020}, but for a prolonged time, and not in the soft state. The observations taken during the 2012 outburst in the pure hard state are much redder compared to all the other data points, especially in the optical CMD representing longer wavelengths ($V$ vs $V$-${i}^{\prime }$) which traces the highest jet contribution. In this case, the significant deviation of the data point away from the disk model can be confidently attributed to the jet, as it starts to dominate the optical emission during the transition to the pure hard state.

\begin{figure*}
\centering
\includegraphics[width=17.7cm,angle=0]{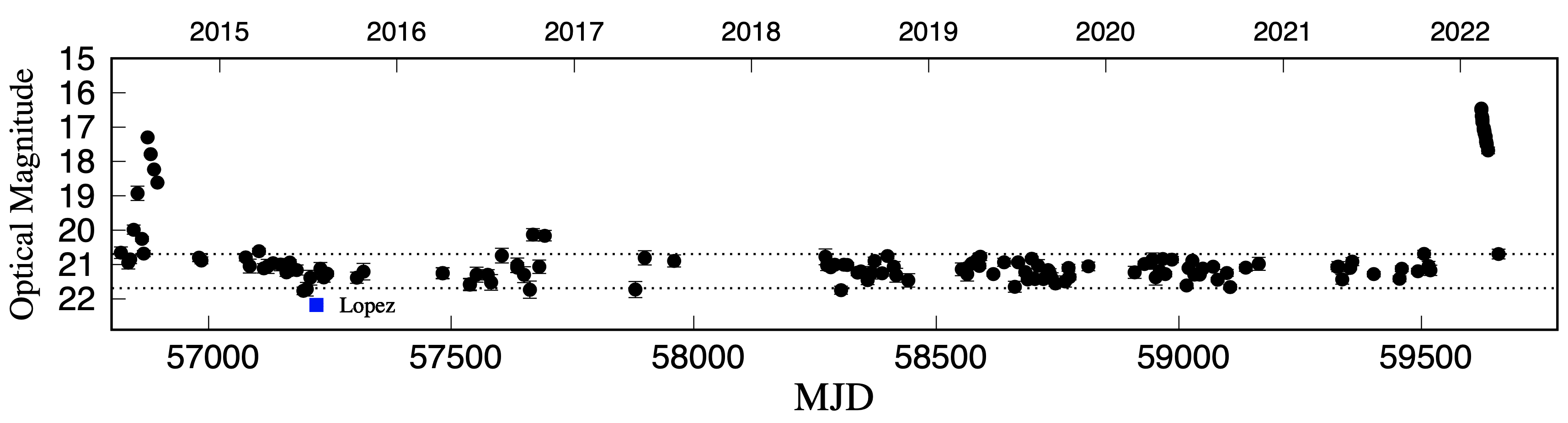}
\caption{Long-term (2014-2022) ${i}^{\prime }$ light curve of J1910.2 from LCO.  The dotted horizontal lines indicate the typical range of quiescence (${i}^{\prime } \sim$20.7-21.7), although it should be noted that they could be contaminated by a few faint, nearby stars (see text).  For comparison, we plot (in blue square) the WHT+ACAM quiescent photometry of \cite{Lopez2019}.}
\label{fig:lc-quiesc-new}
\end{figure*}

Even for the optical/UV CMD, where we plot the bluer wavelengths ($uvm2$ vs $uvm2$-$V$) that is generally dominated by the disk with negligible jet contribution, we find that the data in the soft as well as hard--intermediate (HIMS) and hard states diverged from the disk model. This suggests that the deviation of the 2012 data from the blackbody model is not because of a jet contribution, but probably has its origin in the disk.

The 2022 data are also comparatively brighter and/or redder than the disk model. As these data points are taken during the hard state, and the spectral index is too low for a viscous disk, we attribute this deviation to a jet contribution, just as in the case of the pure hard state data of the 2012 outburst. This is also supported by the AMI-LA radio detections, which showed a considerable rise just before the LCO observations were taken \citep{2022radio}.

\subsection{Variability in quiescence} \label{sec:quiescence}

After the end of the mini-outburst in 2014, we continued the LCO monitoring of J1910.2. In Fig. \ref{fig:lc-quiesc-new}, we plot the long-term (2014-2022) optical (${i}^{\prime }$) light curve of J1910.2, and find that it remained in quiescence throughout this interval, but was variable over a range of ${i}^{\prime }$-band$\sim$20.7-21.7. However, we do note that, during quiescence, the LCO magnitudes \citep[mostly with forced multi-aperture photometry by XB-NEWS at the source position;][]{Goodwin2020} include some contaminating flux from two nearby stars (within 2$^{\prime\prime}$ of the source position), with similar brightness as J1910.2. 

\citealt[][]{Lopez2019} detected J1910.2 at ${i}^{\prime }$ = 22.18$\pm$0.04 on 2015 July 19 (MJD 57222) using the William Herschel Telescope with the auxiliary port camera (ACAM). This is the only other published quiescent magnitude of the source at optical wavelengths, in which the magnitudes obtained are not contaminated by the neighbouring stars. From the finding chart of \cite{Lopez2019}, we know that the two neighboring stars have brightness comparable to J1910.2. Assuming that the ACAM magnitudes are representative of the average magnitude of J1910.2 in quiescence, and that the two neighboring stars are of the same magnitude, we speculate that having all the three stars in the same aperture (as should be the case for LCO data) would give us a flux which is thrice of the real flux. This translates to an optical magnitude of ${i}^{\prime }\sim$21.01$\pm$0.10. In fact, we find that the average quiescent LCO mags is comparable to this, with ${i}^{\prime }$=21.16$\pm$0.29. As we do not have an LCO detection of the source on the same date, a direct comparison of the magnitudes is not possible, but the closest observation with LCO (MJD 57230) is also $\sim$1 mag brighter. As shown in Fig. 8, the long-term LCO light curve suggests accretion variability in quiescence with a range of quiescent magnitudes that are $\sim$0.4 to $\sim$2.0 mag brighter in LCO compared to the \cite{Lopez2019} value (a range that could either be due to varying seeing conditions, and/or due to intrinsic accretion variability). As the amount of flux in the blend depends on the seeing conditions, we cannot completely trust the variability observed. However, we note that although some quiescent variability is expected due to fluctuating seeing, we cannot rule out intrinsic variability, as is seen in many other BHXBs \citep[e.g.][]{koljonen,wu,RussellAlQasim2018}.

\section{Interpretation of the reflares} \label{sec:otherLMXBs}

In many BHXBs, weaker secondary re-brightening events after the source has reached quiescence, either in the form of reflares or mini-outbursts, follow the initial outburst \citep[e.g.][]{Chen1997,Tomsick2003,cuneo,MunozDarias2017}. Such re-brightenings are also observed in neutron star X-ray binaries (NSXBs) and dwarf novae (DNe). This suggests that the cause of at least some post-outburst re-brightening events is related to the accretion process and probably the companion star, and is independent of the nature of the compact object.

Reflares are common in many subclasses of DNe systems; either observed as well-separated rebrightenings after the end of the primary outburst \citep[e.g. V585 Kyr,][]{katoosaki}, or during the decline from the peak of the main outburst caused by a reflection of the cooling wave that propagates from the outer disc edge \citep{Dubus2001,Hameurynew}. The RZ LMi type DNe show fast re-brightenings with very short supercycles \citep[$\sim 20$ days,][]{osaki95}.  These short recurrence times are morphologically similar to the mini-outbursts observed in BHXBs \citep{Hameury,Zhang2019}. The WZ Sge systems, which are an extreme sub-class of SU UMa-type DNe, exhibit much rarer superoutbursts (i.e. very long supercycle times), and are followed by rebrightenings (or reflares) as they return to quiescence \citep{kato}. \cite{Hameurynew} successfully explained the reflares observed in these WZ Sge systems (on the basis of the optical light-curve of TCP J21040470+4631129) using the disk instability model \citep[DIM, e.g.][]{Lasota2001}.

In the framework of DIM, outbursts are thought to be triggered when matter accumulates in the accretion disk during quiescence, thereby heating up the disk and causing the hydrogen in the disk to ionize. This gives rises to a thermal-viscous instability, which initiates the outburst. The DIM predicts the accretion disk to have a minimum amount of matter left at the end of an outburst, and hence cannot easily explain the re-brightening events, because they require a large amount of matter to be left in the disk after an outburst \citep[e.g.][]{Patruno2016_1808}, except under specific conditions \citep[e.g.][]{Zhang2019}. Several other models have been used to explain the mini-outbursts and/or reflares in various compact sources. For example;

\begin{enumerate}
    \item The disk instability model with specific conditions such as the presence of a hot inner disk at the end of the initial outburst \citep[e.g.][]{Zhang2019};
    \item The mass reservoir model as long as the effective viscosity of the disk remains large through the entire sequence of reflares \citep{osaki};
    \item Irradiation of the companion star causing enhanced mass-transfer through X-ray heating \citep[e.g.][]{Hameury};
    \item A smaller discrete accretion event \citep[e.g.][]{sturner};
    \item A small-scale outburst generated by either disk instability or a change in the disk density \citep{Patruno2016};
    \item Enhanced viscosity caused when the outer part of the disk gets irradiated and the generated mass front propagates inward \citep[e.g.][]{Shahbaz1998};
    \item Jet brightening during hard state decay \citep[e.g.][]{Jain2001,saikia2019};
    \item The activation/deactivation of the propeller effect changing the mass accretion rate due to the rapidly rotating NS magnetosphere \citep{Hartman2011,Patruno2016}.
\end{enumerate}

\begin{table*}
\caption{Sample of LMXB outbursts with re-brightenings within 1 year of the last detection of the initial outburst.}
\vspace{-0.3cm}
\label{tab_rebrightenings}      
\centering                       
\begin{tabular}{l c l l r r l r l c}       
\hline    
Source & BH/NS$^1$ & Year$^2$ & Classification$^3$ & $t_{\rm flaring}$ (d)$^4$ & $N_{\rm flares}$$^5$ & Band$^6$ & $\Delta t_{\rm peaks}$$^7$ & State$^8$ & Refs.$^9$ \\
\hline
A\,0620--00 & BH & 1975 & mini-outburst & $\sim 60$ & 1 & optical & -- & -- & 1 \\
GRO~J0422+32 & BH & 1992 & mini-outbursts & $>271$ & 2 & optical & $\sim 113$ & -- & 2 \\
GRS~1716$-$249 & BH & 1993 & reflares & $>400$ & 5 & X-ray & 50--90 & hard & 3 \\
XTE~J1859$+$226  & BH & 1999 & mini-outbursts? & $\sim 75$ & 3 & optical & 20--30 & hard & 4 \\
XTE~J1650--500 & BH & 2001 & reflares? & $>150$ & $\geq 7$ & X-ray & 14.2 & hard & 5,6 \\
Swift~J1753.5--0127 & BH & 2005 & mini-outburst & $>151$ & 1 & both & -- & hard & 7 \\
IGR~J00291+5934 & NS & 2008 & new outburst & $>49.0$ & 1 & both & -- & hard & 8 \\
XTE~J1752--223 & BH & 2010 & reflare? & -- & 1 & both & -- & hard & 9 \\
MAXI~J1659--152 & BH & 2010 & mini-outburst & $89 \pm 15$ & 1 & both & -- & hard & 10,11 \\
MAXI~J1836-194 & BH & 2011 & reflare? & $\sim 75$ & 1 & both & -- & hard & 12,13 \\
Swift~J1910.2--0546 & BH & 2012 & reflares & $> 245.8$ & $\geq 7$ & optical & 42--49 & hard & 14,15 \\
GRS~1739--278 & BH & 2014 & mini-outbursts & $>150$ & 3 & X-ray & $\sim 62$ & hard,soft & 16 \\
V404~Cyg & BH & 2015 & reflares & $>33$ & $>10$ & both & $<1$ & hard & 17,18 \\
MAXI~J1535--571 & BH & 2017 & reflares & $>165$ & $\geq 5$ & X-ray & $31-32$ & hard,soft & 19,20 \\
MAXI~J1820+070 & BH & 2018 & mini-outbursts & $>474$ & 3 & both & $\sim 177$ & hard & 21--28 \\
MAXI~J1348--630 & BH & 2019 & mini-outbursts & $\sim 280$ & $\geq 3$ & both & $\sim 90$ & hard & 29--34 \\
4U~1543--47 & BH & 2022 & mini-outburst,reflares & $> 240$ & $\geq 5$ & both & 20--30 & hard,soft & 35--37 \\
\hline
\end{tabular} \\
$^1$BH = black hole; NS = neutron star.
$^2$Year of initial outburst.
$^3$Re-brightening classification based on \cite{Zhang2019}.
$^4$Total duration of re-brightening interval after initial outburst.
$^5$Number of reflares during re-brightening interval.
$^6$Wavebands of reported re-brightenings (optical, X-ray or both).
$^7$Reflare recurrence times (when >1 flare recorded).
$^8$Re-brightening X-ray state (if known).
$^9$References:
(1) \cite{Charles1998};
(2) \cite{Callanan1995};
(3) \cite{hej1716};
(4) \cite{zurita}; 
(5) \cite{Tomsick2003};
(6) \cite{Tomsick2004};
(7) \cite{Zhang2019};
(8) \cite{Lewis2010};
(9) \cite{corral1752};
(10) \cite{Homan2013};
(11) \cite{CorralSantana2018};
(12) \cite{12};
(13) \cite{13};
(14) this paper;
(15) \cite{Tomsick2013ATel5063};
(16) \cite{YanYu2017};
(17) \cite{MunozDarias2017};
(18) \cite{Kajava2018};
(19) \cite{Parikh2019};
(20) \cite{cuneo};
(21) \cite{Ulowetz2019ATel12567};
(22) \cite{Bahramian2019ATel12573};
(23) \cite{Baglio2019ATel12596};
(24) \cite{Hambsch2019ATel13014};
(25) \cite{Xu2019ATel13025};
(26) \cite{Adachi2020ATel13502};
(27) \cite{Sasaki2020ATel13530};
(28) \cite{1820new}
(29) \cite{AlYazeedi2019ATel13188};
(30) \cite{Pirbhoy2020};
(31) \cite{Shimomukai2020ATel13459};
(32) \cite{Zhang2020ATel13465};
(33) \cite{Baglio2020ATel13710};
(34) \cite{1348new};
(35) \cite{jwaher};
(36) \cite{1543xray1};
(37) \cite{1543xray2}.
\end{table*}

\subsection{Comparison with reflares in other LMXBs}

We compile a list of all BHXB sources (see Table 3) where significant re-brightening was observed within one year of the last detection of the initial outburst (either after it reached quiescence or after a gap where it is uncertain if it reached quiescence). We do not include the recurrent transients \citep[e.g. GX~339-4,][]{Tetarenko2016WATCHDOG} and multi-peak outbursts \citep[e.g. GRO~J1655-40,][]{Chen1997}. Along with BHXBs \citep[e.g., MAXI~J1535-571 and V404~Cyg;]{Parikh2019,cuneo,MunozDarias2017}, such re-brightening events (at different timescales) after a main outburst are also seen in NSXBs \citep[e.g. IGR~J00291+5934;][]{Lewis2010}, as well as the WZ Sge type dwarf novae \citep[see e.g.][]{kato}. We use the observation-based labelling scheme explained in \cite{Zhang2019} to classify the different re-brightening phenomena in this sample, in which a rebrightening is termed as a `reflare' if the flux did not reach quiescence before the rise in amplitude. If the flux reaches quiescence before the rebrightening, we term it a `mini-outburst', provided that the flux ratio between the rebrightening peak and the primary outburst peak is less than 0.7, and the time separating the start of the quiescent period from the start of the rebrightening is less than the duration of the main outburst. On the other hand, if the duration of the main outburst is shorter, or if the flux ratio between the peak of the rebrightening and the peak of the primary outburst is more than 0.7, we term it as a `new outburst' \citep{Zhang2019}.

As discussed in Section 3.1, the 2013 re-brightenings observed in J1910.2 are reflares, and not mini-outbursts. We find that it is one of the very few systems to display such unusually extreme flaring (with more than 7 optical reflares). In most cases, the number of reflares or mini-outbursts seen during the period of re-brightening is less than 5. The only other BHXBs displaying more than 5 reflares within one year of their outbursts are XTE~J1650$–$500 \citep{Tomsick2003, Tomsick2004}, MAXI~J1535$-$571  \citep{cuneo,Parikh2019} and V404~Cyg \citep{MunozDarias2017,Kajava2018}. 

In MAXI~J1535$-$571, at least four reflaring events were seen after the first outburst, all having an approximately constant interval between reflares of $\sim$31-32 days \citep{cuneo}. However, unlike MAXI~J1535$-$571, where a progressive faintness of the reflares is observed, likely due to an emptying reservoir of mass available for accretion \citep{Parikh2019}, the peak magnitude of the 2013 reflares in J1910.2 stayed almost constant. The MAXI~J1535$-$571 reflares also exhibited state transitions and the hysteresis pattern in the HID, which is generally observed only in the main outbursts of LMXBs \citep[except for the mini-outbursts in GRS~1739$-$278,][]{YanYu2017}. Such a comparison is not possible for J1910.2, as there is only one X-ray detection and a few upper limits available from \emph{Swift}/XRT during the reflaring behaviour. The single X-ray detection is as hard as the 2012 outburst hard state decay, so at least for one date during the reflares we can confirm that the source was in the hard state. Moreover, as transitions are usually at higher luminosities, and these reflares are barely detected by \emph{Swift}/XRT, we argue that all the reflares are probably happening in the hard state. We note that the reflares hysteresis loops observed in MAXI~J1535$-$571 occured at almost 100 times lower luminosities than the peak of the main outburst, with the state transitions occuring at a luminosity L$_X \le$ 7$\times10^{36}$ erg $s^{-1}$ \citep[which is the lowest luminosity at which hard-to-soft transitions have been observed in a BHXB, see][]{cuneo}. However the only X-ray detection of J1910.2 available during the reflares is more than 1000 times lower than the outburst peak, suggesting that the 2013 reflares of J1910.2 are happening in the hard state.

\subsection{Origin of the 2013 reflares}

One important observation from the CMD of J1910.2 (see Section \ref{sec:colors}) is that it repeatedly crosses the temperature needed to ionize/neutralize the hydrogen present in the accretion disk during the re-brightening events. Typically at the end of the outburst, the temperature in the outer disk decreases causing the hydrogen in the disk to recombine, and this sends a cooling wave that propagates inwards \citep{Dubus2001,Lasota2001}. It eventually reaches matter in the inner disk that is so hot it cannot be cooled lower than the recombination temperature, so the cooling wave halts. At the radius where the surface density behind the cooling front becomes high enough, the disk becomes thermally unstable, initiating a new heating front to propagate outwards \citep{Dubus2001}. The CMD of J1910.2 suggests that the repeated 2013 reflares are probably due to back and forth propagation of cooling and heating waves in the disk. 

If the instability causing the reflares is originating at the inner disk and then propagating outwards, then the rise time of the repeated reflares, which estimates the propagation time of the heating front, suggests a viscous timescale of $\sim$6 days. A viscous timescale of $\sim$6 days is also measured from the dip in intensity seen during the 2012 outburst, provided it is also caused by reduction in mass transfer into the inner disk \citep{paper1,Degenaar2014}. A measurement of the disk viscosity parameters from the observed light-curve profile \citep[as done using a Hierarchical Bayesian approach with MCMC fitting after removing flares in][]{Bailey} is difficult in this case due to the lack of good coverage during the decay of the reflares. However, overall the general structure of the reflares follow a pattern of rapid heating and a relatively slower fading, similar to what is observed during main outbursts.

Numerical simulations of the DIM automatically predict reflares which are spontaneously created through repeated heating/cooling waves that cyclically ionize and recombine the accretion disk, although the numerically produced light curves do not generally resemble those observed \citep{Dubus2001,meyer,Hameurynew}. Moreover, the reflares predicted by the DIM require the density of matter to be depleted with each subsequent reflare, and hence a progressive faintness in amplitude is expected \citep{Dubus2001}; which is not observed in the case of J1910.2. However, we speculate that a heated up companion can continuously dump matter in the disk, due to its expansion from being heated by the X-rays of the 2012 outburst. This enhanced mass transfer from the companion (in addition to the steady accretion from the companion that happens all the time) can result in an almost constant amplitude during the reflares. Another possibility is that the X-ray and optical emission shows different things. X-rays trace the mass accretion rate close to the black hole, and a decreasing trend of peak X-ray luminosities is expected \citep[as seen in the X-ray light curves of MAXI~J1535$-$571 reflares]{cuneo}. However, the constant peak optical magnitude could correspond to the position in the CMD where the disk reaches above the H ionization temperature. From the CMDs of J1910.2 (see Fig. 7), we find that the data follow the disk model very well, suggesting that the emitting area is roughly constant during the reflares. Hence it is possible that we are probing different mechanisms in both the wavelengths; we could be looking at a constant area blackbody heating and cooling with optical, while tracing the mass accretion rate in X-rays. This could be another reason why we do not have a decreasing trend of peak optical fluxes as also seen in the optical light curves of GRO~J0422+32 reflares by \cite{Callanan1995}.

Overall, it is not completely clear if the reflares are caused by the same hydrogen ionization instability which triggers the main outburst or has a different origin mechanism. However, from the changes in temperature observed during the reflares (which repeatedly cross the H ionization temperature), we consider the back and forth propagation of heating/cooling waves to be the most likely explanation for the 2013 reflares.

\subsection{Origin of the 2014 mini-outburst}

Large-amplitude optical oscillations or violent reflares seen on shorter timescales (on timescales of hours) in sources like V404 Cyg are expected in long-period systems. The disk in such systems is much larger, and the surface densities in the outer disk will be too low to have sustained mass accretion in the inner disk, which is required for longer-timescale reflares \citep{Kimura2016}. In fact, the longer-timescale 2014 mini-outburst as seen in J1910.2 is expected to be more common in BHXB systems with a shorter orbital period ($<$ 7 hr). In such short-period systems, it is speculated that the outer disk has a high enough temperature for the heating front to remain hot, thereby triggering a mini-outburst \citep{Zhang2019}. Due to the lack of any deep soft X-ray observation during the outburst fade, or before this 2014 mini-outburst, we do not have direct confirmation for the presence of a hot inner disk. However, recent optical fast photometry of J1910.2 indeed suggests an orbital period of $<$ 7.4 hrs (Saikia et al. submitted). Previously, \cite{Casares2012} had reported an orbital period $>$ 6.2 hr from their spectroscopic study, study, assuming that the velocity changes in H$\alpha$ emission are cause by binary motion. Later, a fairly short orbital period ($\sim$ 2–4 hrs) was proposed based on the small size of its disk with a radius of $\sim 4 \times 10^9$ cm \citep{Degenaar2014} and its variable optical emission \citep{llyod}. Such a short orbital period can ensure the presence of a hot inner disk at the end of the outburst decay, which could have triggered the mini-outburst seen in J1910.2. 

\section{Summary and conclusions} \label{sec:conclusions}

In this work, we present long-term optical monitoring of the candidate black hole transient X-ray binary Swift~J1910.2$-$0546 from 2012--2022 using the Faulkes Telescopes and LCO. We report two periods of re-brightening activities previously undocumented in the literature, which include a series of at least seven quasi-periodic, high amplitude ($\sim 3$ magnitudes) optical reflares in 2013, and a mini-outburst with two peaks in 2014. We find that the source shows a bluer when brighter behavior during both the re-brightening episodes of 2013 and 2014. The optical colors during these epochs are consistent with a blackbody heating and cooling between 4500 and 9500 K, suggesting that the flares could be caused by repetitive heating and cooling waves travelling through the accretion disk. We compare them with re-brightening events observed in all the other BHXBs within one year of an outburst, and show that the repeated reflaring behaviour of J1910.2 is highly unusual among BHXBs. We discuss the different scenarios which could cause such extreme flaring, and propose that they arise from a sequence of heating and cooling front reflections in the accretion disk following the disk instability model, probably due to the presence of a hot inner disk at the end of the 2012 outburst. 

\acknowledgments

DMR and DMB acknowledge the support of the NYU Abu Dhabi Research Enhancement Fund under grant RE124. This work uses data from the Faulkes Telescope Project, which is an education partner of Las Cumbres Observatory (LCO). The Faulkes Telescopes are maintained and operated by LCO. This work also makes use of data supplied by the UK \emph{Swift} Science Data Centre at the University of Leicester, and the MAXI data provided by RIKEN, JAXA and the MAXI team.

\clearpage
\restartappendixnumbering

\bibliographystyle{aasjournal}

\end{document}